\documentclass{article}
\usepackage[xdvi]{graphics, rotating}
\usepackage{latexsym}
\usepackage{amsmath}
\usepackage{amsfonts}
\usepackage{amscd}
\usepackage{a4wide}
\def\N0{\mathbb N_0}
\def\v{v}

\newtheorem{lemma}{Lemma}

\begin{document}
\title{Formal derivation of an exact series expansion
for the Principal Field Emission Elliptic Function $\v$}

\author{Jonathan H.B. Deane$^1$, Richard G. Forbes$^2$ and R.W. Shail$^1$\\ \\
$^1$Department of Mathematics,\\
$^2$Advanced Technology Institute (BB),\\
University of Surrey, Guildford GU2 7XH, UK\\
\\
E-mail: J.Deane@surrey.ac.uk, r.forbes@ieee.org}

\date{7 August 2007}

\maketitle

\begin{abstract}
\noindent
An exact series expansion is now known for the Principal Field Emission Elliptic Function $\v$, 
in terms of a complementary elliptic variable $l'$ equal to $y^2$, where $y$ is the 
Nordheim parameter. This expansion was originally found by using the algebraic 
manipulation package MAPLE. This paper presents a formal mathematical derivation.
Recently, it has been discovered that $\v(l')$ is 
a particular solution of the ordinary differential equation (ODE) $l'(1-l')$d$^2\v$/d$l'^2=n\v$, 
when the index $n = 3/16$. This ODE appears to be new in mathematical physics and elliptic-function 
theory. The paper first 
uses an 1876 result from Cayley to establish the boundary condition that d$\v$/d$l'$ satisfies 
as $l'$ tends to zero.  It then uses the method of Frobenius to obtain two linearly independent 
series solutions for the ODE, and hence derives the series expansion for $\v(l')$. It is 
shown that terms in $\ln{l'}$ are required in a mathematically correct solution, but fractional 
powers of $l'$ are not. The form of the ODE also implies that simple Taylor expansion methods
cannot generate good approximation formulae valid over the whole range $0 \leq l' \leq 1$;
this conclusion
may also apply to barriers of other shapes. It is hoped that this derivation might serve as 
a paradigm for the treatment of other tunnelling barrier models for cold field electron emission, if 
in any particular case an ODE can be found for which the tunnelling-exponent correction function is 
a particular solution.     
\end{abstract}

\section{Introduction}

Fowler-Nordheim (FN) tunnelling~\cite{FN28} is electric-field-induced electron tunnelling through a 
roughly triangular barrier. When the emission barrier is strong and the penetration coefficient 
small, there is a low-temperature emission regime (including room temperature) known as 
``cold field electron emission (CFE)". Tunnelling and CFE are processes of significant 
technological interest, in particular for the prevention of vacuum breakdown, the development 
of cold-cathode electron sources, and internal electron transfer processes in some types 
of  electronic device. Tunnelling and CFE are at the heart of several multi-billion-dollar industries.  

CFE from a metal-like conduction band is described by a family
of approximate equations known as ``Fowler-Nordheim-type" (FN-type)
equations. FN-type equations for the emission-current density $J$, as a
function of the local work-function $\phi$ and the local barrier field $F$,
have the form
\begin{equation}
J = \lambda a \phi^{-1}F^2\exp(-\mu b\phi^{3/2}/F)
\label{FNgen}
\end{equation}
where $a$ and $b$ are the First and Second FN Constants as usually defined
(e.g. \cite{F04}), and $\lambda$ and $\mu$ are generalised correction factors. The
forms of $\lambda$ and $\mu$ depend on the assumptions and approximations made
in the derivation, in particular on the shape assumed for the tunnelling barrier.  

For a planar emitter, a ``Schottky-Nordheim" (image-rounded) barrier model~\cite{S14,N28} is 
normally used. In this case, $\mu$ is given by the mathematical parameter $v_{\rm{F}}$, which 
is a particular value (see below) of the Principal Field Emission Elliptic Function $\v$. 
This function $\v$ is obtained by applying the simple-JWKB approximation~\cite{Jeff,FF} to 
a Schottky-Nordheim (SN) barrier, and can be defined by a JWKB integral (see, for example, 
~\cite{prsa}). Although always called an ``elliptic function", $\v$ is better described as a special kind of complete elliptic integral. An expression for $\v$ in terms of the complete
elliptic integrals of the first and second kinds, 
$K(m)$ and $E(m)$, where $m$ is the elliptic parameter as used in~\cite{AS}, has been known 
for many years, and numerical values have long been available~\cite{BKH}.

Recently,~\cite{prsa,APL}, an exact series expansion (see below) has been found for $\v$, 
in terms of the elliptic variable $l'$ defined by
\begin{equation}
l'  =  [(1 - m)/(1+m)]^2.
\label{ldef}
\end{equation}
It follows that
\begin{equation}
m(l') = (1-\sqrt{l'})/(1+\sqrt{l'}).
\label{lmdef}
\end{equation}
The primed symbol $l'$ was chosen, in accordance with a normal convention in elliptic-function theory, 
because $l'$ is a ``complementary" elliptic variable in the sense that $l' \rightarrow 0$ as 
$m \rightarrow 1$. Knowing the form of the exact expansion for $\v(l')$  has enabled us to 
develop~\cite{prsa} numerical approximation formulae for $\v(l')$ and d$\v$/d$l'$, with 
absolute error $|\epsilon| <7\times 10^{-10}$, that substantially outperform earlier 
numerical approximations of equivalent complexity~\cite{Hast}. It also provides a 
mathematical explanation for the success of the simple approximation formula~\cite{prsa,APL}
\begin{equation}
\v(l')  \approx  1 - l' + (1/6)\,l'\ln l',
\label{3-term}
\end{equation}
which has the merits that it is exact at $l' = 0$ and $l' = 1$, and (when assessed over the whole range 
$0 \leq l' \leq 1$) has absolute error $|\epsilon| < 0.0025$ and outperforms all existing formulae of 
equivalent complexity~\cite{prsa}.

In CFE, the physical parameter $f$ (the
``scaled barrier field") is defined by $f=F/F_\phi$, where $F_\phi$ is the barrier
field necessary to reduce to zero a barrier of zero-field height $\phi$.
The values of $F$ and $f$ determine the strength of the tunnelling barrier.
For the SN barrier, this physical parameter $f$ can be identified with the mathematical parameter $l'$, and $\mu$ in eq.~(\ref{FNgen}) is given by $\v(l'=f)$.

The complete elliptic integrals of the first and second kinds may be defined~\cite{AS} in terms of 
the elliptic parameter $m$ by 
\begin{equation}
K(m) = \int_0^{\pi/2}\frac{\mathrm{d}\phi}{\sqrt{1-m\sin^2\phi}}
\mbox{\hskip0.4in and\hskip 0.4in}
E(m) = \int_0^{\pi/2}\sqrt{1-m\sin^2\phi}\, \mathrm{d}\phi.
\end{equation}
Using~(\ref{lmdef}), $K$ and $E$ can, in principle, be expressed in terms of $l'$ rather than $m$. 
The expression for $\v(l')$ in terms of $K$ and $E$ then is~\cite{prsa}:
\begin{equation}
\v(l')  =  (1+\sqrt{l'})^{1/2}\, [E(m(l')) - \sqrt{l'}K(m(l'))]
\label{vEK}
\end{equation}
where $m(l')$ is given by equation~(\ref{lmdef}).

The exact series expansion for $\v(l')$ was originally generated~\cite{APL} by the
algebraic manipulation package MAPLE, by expanding an expression equivalent
to~(\ref{vEK}).  Numerical values obtained from MAPLE agreed with values
obtained by numerical evaluation of the relevant JWKB integral, to better
than 12 decimal places, strongly indicating that both methods were
mathematically sound. We subsequently found algebraic formulae that
reproduced the MAPLE result (see~\cite{prsa}, Appendix~1). However,
``manual" derivation of higher-order terms in $\v(l')$ is excessively
laborious;  also, this method does not bring out the underlying
mathematics.

Seeking a better mathematical derivation, we found~\cite{prsa} that
$\v(l')$ is a particular solution of the ordinary differential equation (ODE)
\begin{equation}
l' (1 - l')\;\frac{\mbox{d}^2W}{\mbox{d}l'^2} = nW,
\label{supeq}
\end{equation}
when $n = 3/16$ and certain boundary conditions (see below) are
satisfied. Equation~(\ref{supeq}) appears to be new to mathematical physics
and elliptic-function theory. Its solution has not been found in easily available
mathematical literature, but does not involve any new mathematical principles.
Applying the 
method of Frobenius~\cite{wylie} generates a series expansion with coefficients 
defined by recurrence relations (see below). These formulae correctly generate 
the numerical coefficients already reported~\cite{APL}.

This finding has itself been reported~\cite{prsa}, but the full proof was too lengthy
to include in earlier work. This paper presents the full mathematical derivation of
the relevant recurrence relations. In part, the aim is to provide closure on
the mathematics of the (simple-JWKB treatment of the) SN barrier; but 
we hope this method might also be a paradigm for finding exact solutions for 
correction factors for barriers of other shapes. It would be applicable where 
a method can be found for converting the JWKB-type integral definition of the 
correction function into an equivalent differential equation.

The structure of the paper is as follows. Section 2 derives boundary conditions that
$\v$ must satisfy, Section 3 derives general solutions for~(\ref{supeq}), 
Section 4 uses these to derive a series expansion for $\v$, and Section 5 provides discussion.  

\section{Boundary conditions}

To establish the expansion for $\v$, we need the boundary conditions that $\v(l')$ and 
d$\v$/d$l'$ satisfy at $l'=0$. It is well known in CFE theory that $\v(0)=1$. 
However, d$\v$/d$l'$ becomes infinite as $l'\rightarrow 0$, so the boundary condition 
on d$\v$/d$l'$ has to take the slightly unusual form that ``d$\v$/d$l'$ becomes infinite 
in the correct way". We develop both conditions from a result proved by Cayley in his 
1876 textbook~\cite{cayley}.  This section is presented in the form of three lemmas.

\begin{lemma}
As $l'$ approaches zero from above, the function $K(m(l'))$ is given by
\begin{equation}
K(m(l')) = (3/2)\ln2 - (1/4)\ln{l'} + O(\sqrt{l'}).
\label{Kapp}
\end{equation}
\end{lemma}

{\bf Proof}.  From~(\ref{lmdef}), we have
  
$$(1-m) =  {\frac{2\sqrt{l'}}{1+\sqrt{l'}}} .$$
Formula 17.3.26 in~\cite{AS} is derived from Cayley's result and  states that

$$\lim_{m\rightarrow 1} K(m) = \ln{[4/\sqrt{(1-m)}]}.$$
Hence
$$\lim_{l'\rightarrow 0} K(m(l'))
= \lim_{l'\rightarrow 0} \ln \left[ {4\sqrt{\frac{1+\sqrt{l'}}{2\sqrt{l'}}}}\right]
= \lim_{l'\rightarrow 0}\left[(3/2)\ln 2 - (1/4)  \ln{l'} + O(\sqrt{l'})\right],$$
as required. We have thus proved a Cayley-type result for $K$ as a
function of $l'$.  This result  is key to deriving the series expansion for $\v$. 

\begin{lemma}
$\v(0)=1.$
\end{lemma}

{\bf Proof}. Using lemma 1, we find that, in the limit of small $l'$, the term 
involving $K(m(l'))$ in definition~(\ref{vEK})  takes the form

$$-\lim_{l'\rightarrow 0} \left[{(1+\sqrt{l'})^{1/2} 
\sqrt{l'} [(3/2)\ln2 - (1/4)\ln{l'}]} \right] = 0.$$

When $l'=0$, then $m=1$, and the term in $E(m(l'))$ in definition~(\ref{vEK}) reduces to

$$E(m=1) = \int_0^{\pi/2}\cos\phi\, d\phi = 1.$$
It follows that $\v(0)=1$.  This is a well-known result in CFE theory, but for 
completeness we have given formal proof here.

\begin{lemma}
\begin{equation}
\mbox{\hskip0.4in}
\lim_{l'\rightarrow 0} \left[\mathrm{d}\v/\mathrm{d}l' - (3/16)\ln l' \right]
   = -(9/8)\ln 2
\label{limdvdl}
\end{equation}
\label{limits}
\end{lemma}

{\bf Proof}. From~\cite{prsa} we have the result that
\begin{equation}
\frac{\mathrm{d}\v}{\mathrm{d}l'} = -\frac{3\,K(m(l'))}{4 \left(1 + \sqrt{l'}\right)^{1/2}}.
\label{dvdl}
\end{equation} 
Using~(\ref{Kapp}), we have, for small $l'$, that

$$\frac{\mathrm{d}\v}{\mathrm{d}l'} \approx -\frac{3}{4} \left[{1-\frac{\sqrt{l'}}{2}}\right]
\left[{\frac{3}{2}\ln{2} - \frac{1}{4} \ln{l'}}\right].$$
The limiting form for d$\v$/d$l'$ in~(\ref{limdvdl}) follows.

\section{General solution of the ODE}
We now solve the ODE. Mathematically, it is convenient to solve the general equation, 
i.e.~(\ref{supeq}), and then put $n$ equal to 3/16 .  We use the symbol $W$ to 
denote a general solution of the relevant equation, whether the index is taken 
generally as $n$ or specifically as 3/16.

We use the method of Frobenius~\cite{wylie,Birk} to find a series expansion for 
$W(l')$. Initially assume a power series solution of the form 
$l'^r \sum_{i=0}^{\infty}g_i l'^i$, where $r$ and the $g_i$ are to be determined. 
Substituting into~(\ref{supeq}) and equating powers of $l'$, we find
\begin{equation}
g_0r(r-1)l'^{r-1}+l'^r\sum_{i=0}^\infty{l'^i[-(i+r-1)(i+r)g_i - ng_i + (i+r)(i+r+1)g_{i+1}]} = 0.
\label{indic}
\end{equation}
Since this is an identity for all powers of $l'$, the first term has to be identically zero and 
there is a recurrence relation for the coefficient $g_i$:
\begin{equation}
g_{i+1} = \frac{(i+r-1)(i+r)+n}{(i+r)(i+r+1)}g_i, \;\;\; i\geq 0.
\label{gi}
\end{equation}

Obviously, if $g_0=0$ then all $g_i$ ($i\geq 1$) are also zero, and $W$ is identically equal 
to 0; so the $g_0=0$ case cannot contribute to the general solution. Hence, we 
deduce that the indicial equation is $r(r-1) = 0$. This equation has the two 
roots $r_1=1, r_2=0$. Since $|r_1 - r_2|$ is an integer, two linearly independent 
solutions $W_{\rm{A}}, W_{\rm{C}}$ are~\cite{wylie,Birk}:

\begin{equation}
W_{\rm{A}}(l') = \alpha W_1(l'),
\label{indA}
\end{equation}

\begin{equation}
W_{\rm{C}}(l') = \chi [W_2(l') +W_1(l') \ln{l'}],
\label{indC}
\end{equation}
where $\alpha$ and $\chi$ are arbitrary constants. We take $\alpha$ equal to unity, without 
loss of generality, but leave $\chi$ to be chosen appropriately later. $W_1(l')$ 
and $W_2(l')$ are the series expansions that correspond to the values $r=1, r=0$, respectively, and are written in the forms
\begin{equation}
W_1(l') = l' \sum_{i=0}^\infty a_i l'^i,\;\;\; W_2(l') = \sum_{i=0}^\infty b_i l'^i,
\label{ser}
\end{equation}
where the $a_i$ and $b_i$ are coefficients to be determined.

The general solution of~(\ref{supeq}) can thus be put in the form
\begin{equation}
W(l') = A W_{\rm{A}}(l') + C W_{\rm{C}}(l')
\label{gen}
\end{equation}
where $A$ and $C$ are arbitrary constants to be determined by the boundary conditions, and 
the two independent solutions have the forms
\begin{equation}
W_{\rm{A}}(l') = \sum_{i=0}^\infty a_i l'^{i+1}.
\label{VA}
\end{equation}
\begin{equation}
W_{\rm{C}}(l') = \chi \left[{\sum_{i=0}^\infty b_i l'^i + 
l'\ln{l'}\sum_{i=0}^\infty a_i l'^i}\right].
\label{WC2}
\end{equation}
Note that the multiplier $l'\ln{l'}$ is an intrinsic part of 
the correct mathematical solution of~(\ref{supeq}).

The coefficients $a_i$ have the properties of the coefficients $g_i$ when $r$ is 
taken as 1 in~(\ref{indic}). So, from~(\ref{gi}):
\begin{equation}
a_{i+1} = \frac{i(i+1)+n}{(i+1)(i+2)}\;a_i, \;\;\; i\geq 0.
\label{ai}
\end{equation}
As far as the solution $W_{\rm{A}}$ is concerned, the value of $a_0$ can be 
chosen arbitrarily without loss of generality, because all values of 
$a_i\;(i\geq 1)$ are proportional to $a_0$.
 
To determine the coefficients $b_i$, we substitute form~(\ref{WC2}) into~(\ref{supeq}), eliminate the terms in ln$l'$ by noting that $W_1$ is a 
solution [hence, $l'(1-l')$d$^2W_1$/d $l'^2 - nW_1 =0$], expand $W_2(l')$ in series 
form, and eventually obtain
 
\begin{equation}
 n (b_0 + b_1 l') - 2b_2 l'
+ \sum_{i=0}^\infty\left([\{(i+1)(i+2) + n \}b_{i+2} - (i+2)(i+3)b_{i+3}\right]l'^{i+2}) 
+\sum_{i=0}^\infty c_i l'^i  = 0.
\label{id1}
\end{equation}
where
\begin{equation}
\sum_{i=0}^\infty c_i l'^i = 2(l'-1)\mathrm{d}W_1/\mathrm{d}l' - (1-1/l') W_1\;\;
=\;\;(l'-1)\sum_{i=0}^\infty (2i+1)a_i l'^i.
\label{id2}
\end{equation}
Hence
\begin{equation}
c_0 = -a_0 
\mbox{\hskip0.4in and\hskip 0.4in}
c_i = (2i-1)a_{i-1} - (2i+1)a_i, \;\;\; i \geq 1.
\label{ci}
\end{equation}

In order for~(\ref{id1}) to be an identity, we need the coefficients of 
$l'^0$ and $l'^1$ to be zero, which requires that: $c_0 + nb_0 = 0$; and $c_1 + nb_1 - 2b_2 = 0.$
This yields
\begin{equation}
b_0 = -c_0/n = a_0/n,
\label{b0}
\end{equation}
\begin{equation}
b_2 = (c_1+nb_1)/2 = (a_0-3a_1+nb_1)/2. 
\label{b2}
\end{equation}
More generally, the requirement that in~(\ref{id1}) the coefficient of $l'^i$ be zero yields

\begin{equation}
\{(i+1)(i+2) + n \}b_{i+2} - (i+2)(i+3)b_{i+3} + c_{i+2} =0,
\;\;i\geq 0,
\label{bi}
\end{equation}
which on re-arrangement yields

\begin{equation}
b_{i+1} = \frac{c_{i}+\{(i-1)i + n\}b_{i}}{i(i+1)}, \;\;i\geq 2.
\label{bi1}
\end{equation}
Equation~(\ref{bi1}) also gives $b_2$ correctly;  so, using~(\ref{ci}), we can write

\begin{equation}
b_{i+1} = \frac{(2i-1)a_{i-1} - (2i+1)a_{i} +\{(i-1)i + n\}b_{i}}{i(i+1)}\;, \;\;i\geq 1.
\label{bi2}
\end{equation}

In summary, the value of $b_0$ is fixed by~(\ref{b0}), and the values of 
$b_i\;(i\geq 2)$ are determined, via~(\ref{bi2}), by the choice of $b_1$.

If $W_{\rm{C}}$ is to be a linearly independent solution of~(\ref{supeq}), then 
any multiple 
of  $W_{\rm{C}}$ must also be a solution. We can satisfy this requirement by ensuring 
that each of the coefficients $a_i, b_i$ is individually proportional to $a_0$. This 
is already true for all the $a_i$ and for $b_0$. It is sufficient that we further put
\begin{equation}
b_1 = \beta a_0,
\label{b1}
\end{equation}
where $\beta$ is a constant to be chosen;  from~(\ref{bi2}), this ensures
that all $b_i\;(i\geq 2)$ are proportional to $a_0$.

At this point, we have derived general forms for two linearly 
independent solutions of~(\ref{supeq}). To derive specific forms for these 
independent solutions, we need to choose values for $a_0$, $\beta$ and $\chi$. 
In principle, these might be chosen in various ways. We choose them here in such a 
way as to make the boundary conditions at $l'=0$ straightforward to apply.

From equations~(\ref{gen}) to~(\ref{WC2}), and~(\ref{b0}), the form for $W(l')$, 
up to terms involving $l'$, is:
\begin{equation}
W(l') = Ca_0 \chi /n + (Aa_0+C\chi \beta a_0 )l' + C \chi a_0l' \ln{l'} + \ldots ,
\label{W}
\end{equation}
and the corresponding form for d$W$/d$l'$ is
\begin{equation}
\mathrm{d}W/\mathrm{d}l' = Aa_0 + C \beta \chi a_0 + C \chi a_0 + C \chi a_0 \ln{l'} +  \ldots
\label{dW/dl'}
\end{equation}
From~(\ref{W}), when $l'=0$:
\begin{equation}
C = (n/\chi)(W_0/a_0).
\label{C}
\end{equation}
where $W_0$=$W(0)$. Clearly, a convenient way to deal with~(\ref{dW/dl'}) is to 
choose $\beta$=$-1$.  Equations~(\ref{W}) and~(\ref{dW/dl'}) are now reduced to
\begin{equation}
W(l') = W_0 + [Aa_0-nW_0]l' + nW_0l' \ln{l'} + \ldots
\label{W2}
\end{equation}
\begin{equation}
\mathrm{d}W/\mathrm{d}l' = Aa_0 + nW_0 \ln{l'} +  \ldots
\label{dW/dl'2}
\end{equation}
Since these equations contain $a_0$ only in the combination $Aa_0$, we can take 
$a_0=1$ without loss of generality; then $b_0=1/n$ and $b_1=-1$. Values of 
$a_i\;(i \geq 1)$ and $b_i  \; (i \geq 2)$ can then be obtained from 
the recurrence formulae above.

At this point, we have the situation that $W_{\rm{A}}(0)=0$ and
\begin{equation}
W_{\rm{C}}(0)=\chi b_0=\chi/n
\label{WC0}
\end{equation}
It is convenient to have $W_{\rm{C}}(0)=1$, so we choose $\chi=n$, which also 
simplifies~(\ref{C}) to
\begin{equation}
C = W(0)/a_0=W(0).
\label{C2}
\end{equation}
This completes the definition of two convenient, linearly independent solutions of
ODE~(\ref{supeq}), and enables the general solution to be written in the form

\begin{equation}
W(l') = 1 +  \sum_{i=0}^\infty (Aa_i +Cnb_{i+1} +Cna_i \ln{l'})l'^{i+1},
\label{result}
\end{equation}
where $a_0=1, b_1=-1$, and the other coefficients are given by 
the recurrence relations~(\ref{ai}) and~(\ref{bi}).  Note that this form (and the value derived below for $C$) differ slightly from a form proposed earlier~\cite{Chicago}, because we have defined $W_{\rm{C}}$ in a slightly different (more convenient) way here.

\section{Series expansion for function $\v$}

At this point, we revert to discussion of the ODE of index 3/16, and derive the
expansion for $\v(l')$. For $n=3/16$, 
the first few terms of the two independent  solutions are
\begin{equation}
 W_{\rm{A}}(l') = l' + \frac{3}{32}l'^2 + \frac{35}{1024}l'^3 + \ldots,  
\end{equation}
\begin{equation}
 W_{\rm{C}}(l') = 1 -\frac{3}{16}l' + \frac{51}{1024}l'^2 + \frac{177}{8192}l'^3 +
 \ldots + \left(\frac{3}{16} + \frac{9}{512}l' + \frac{105}{16384}l'^2 + \ldots\right) l'\ln{l'}.  
\end{equation}

Using the boundary condition $W_0 = \v(0) = 1$,~(\ref{C2}) yields $C=1$. For the 
second boundary condition we use (as noted earlier) the slightly unusual requirement 
that, as $l'$ tends to zero, the derivative d$W$/d$l'$  ``tends to infinity in the 
correct way".    Comparing~(\ref{limdvdl}) and~(\ref{dW/dl'2}) yields the 
following identity, valid in the limit of small $l'$-values:

\begin{equation}
(3/16) \ln{l'} - (9/8) \ln2 = nW_0 \ln{l'} + Aa_0 
\end{equation}
Since $n=3/16$ and $W_0=1$, the terms in $\ln{l'}$ cancel.  So, since $a_0=1$:
\begin{equation}
A = -(9/8) \ln2.
\end{equation}

So the first few terms of the expansion for $v(l')$ are:
$$\v(l') = 1 -\left(\frac{9}{8}\ln 2+\frac{3}{16}\right)l' - \left(\frac{27}{256}\ln 2-\frac{51}{1024}\right)l'^2
-\left(\frac{315}{8192}\ln2 - \frac{177}{8192}\right)l'^3+ \ldots$$
\begin{equation}
+l'\ln l' \left(\frac{3}{16} + \frac{9}{512}l' + \frac{105}{16384}l'^2 + \ldots\right).
\label{valg}
\end{equation}
This series has the form reported earlier~\cite{prsa,APL}, found by using MAPLE.  Evaluating the coefficients to 5 decimal places yields
\begin{equation}
\v(l') = 1-0.96729l'-0.02330l'^2-0.00505l'^3+ \ldots + l' \ln l'(0.18750+0.01758l'+0.00641l'^2+ \ldots).
\end{equation}
By dividing the first expression by $1-l'$, this can be put  into a form that is exact at both $l'=0$ 
and $l'=1$, even when the two series expansions below are truncated:
\begin{equation}
\v(l') = (1-l')(1+0.03271l'+ 0.00941l'^2+ \ldots) + l'\ln{l'}(0.18750+0.01758l'+0.00641l'^2+ \ldots).
\label{vfnum2}
\end{equation}
Both of the series in~(\ref{vfnum2}) have good convergence properties in the range 
$0 \leq l' \leq 1$, so this form is useful for numerical approximations~\cite{prsa}.

\section{Discussion}

For most of the last 50 years, the correction function $\v$ has been expressed in CFE theory 
as a function of the Nordheim parameter $y=\sqrt{l'}$~\cite{N28}, rather than $l'$. 
In~\cite{APL} it was argued that, because the exact series expansion~(\ref{valg}) contains no 
terms in $\sqrt{l'}$, it is mathematically more natural to use $l'$ as the independent 
variable.  The derivation here, using the method of Frobenius, confirms this formally.  
Terms in $\ln{l'}$ are an intrinsic part of the correct mathematical solution, but 
fractional powers of $l'$ are not necessary.

Approximate formulae such as~(\ref{3-term}) cannot be derived by simple 
Taylor expansion methods.  It is implicit in the mathematical analysis here that
simple Taylor expansion methods cannot generate good approximate formulae
valid for the whole range $0 \leq l' \leq1$. This conclusion may also be
applicable to tunnelling barriers of other shapes.

Equation~(\ref{supeq}), the defining ODE, can be transformed to be in terms of the variable $y$. This yields
\begin{equation}
(1-y^2)\frac{d^2W}{dy^2} +\frac{1-y^2}{y}\frac{dW}{dy} - \frac{3}{4}W = 0.
\label{dey}
\end{equation}
Obviously, this equation is more complex than~(\ref{supeq}). This is a further good 
mathematical reason for choosing to use the variable $l'$, rather than $y$, in CFE theory.  As discussed elsewhere~\cite{prsa}, an important implication of the role
of $l'$ in the mathematics is that the natural variable to use in physical discussions of CFE theory is the scaled barrier field $f$.

The algebraic manipulation package MAPLE has played a crucial role in stimulating
this work, because our use of MAPLE~\cite{APL} drew attention to the existence and form of the series expansion for $\v(l')$ and provided a result to aim for. Derivation
of this series
by inserting series expansions for $K(l')$ and $E(l')$ into~(\ref{vEK}) proved
excessively laborious if performed by hand~\cite{prsa}, even when we had found
the Cayley forms~\cite{cayley} for the
series expansions of $K(m)$ and $E(m)$, (which, interestingly, are not reported in
several of the elliptic-function handbooks consulted). So we looked for the ODE that
$v(l')$ obeyed and found~(\ref{supeq})~\cite{prsa}.   

Luckily, the ODE~(\ref{supeq}) has a very simple form. Nevertheless,  the 
derivation of solutions is quite a lengthy mathematical process. Wylie~\cite{wylie} 
(p. 322) describes the solution of equations of this general kind as ``straightforward but 
very tedious"; we confirm his description. Much of Section 3 here could, in principle,
have been written in the late 1800s, or at any subsequent time, but there has been
little incentive to do so.

Overall, it is not entirely surprising that getting a series expansion/definition
for $\v$ in place has taken nearly 80 years, measured from the original (incorrect) attempt by Nordheim~\cite{N28} to derive an exponent correction function for the
SN barrier.  Essentially, it is the relatively recent introduction of reliable
computer algebra packages that has made the discovery of this expansion (and its
mathematical origin) feasible, rather than so difficult as to be unlikely to happen.  The problem for CFE and the SN barrier (even when using the simple-JWKB approximation) has been that the necessary mathematical functions have not been solidly in place
when physicists needed them, in the 1920s and 1950s; one might contrast this with the wave-mechanical
analysis of the hydrogen atom in the 1910s, which needed mathematical functions that had been well defined for very many years. 

As indicated in the introduction, our hope is that the derivation in this paper may be able 
to serve as a paradigm for the treatment of other barrier models, particularly for realistic 
models for the potential-energy variation above sharply curved emitters. The keys, in each case, will be to find an ODE that the tunnelling-exponent correction function satisfies, and a suitable formulation of the boundary conditions.


\begin{thebibliography}{10}
\bibitem{FN28} Fowler R H and Nordheim L W 1928 Electron emission in intense electric fields. {\it Proc. R. Soc. Lond. Ser. A} {\bf 119} 173-181.
\bibitem{F04} Forbes R G 2004 Use of energy-space diagrams in free-electron models of field electron emission {\it Surf. Interface Anal.} {\bf 36} 395-401.
\bibitem{S14} Schottky W 1914 \"{U}ber den Einfluss von Strukturwirkungen, besonders der Thomsonschen Bildkraft, auf die Elektronenemission der Metalle. {\it Physik. Zeitschr.} {\bf 15,} 872-878.
\bibitem{N28} Nordheim L W.1928 The effect of the image force on the emission and reflexion of electrons by metals. {\it Proc. R. Soc. Lond. Ser. A.} {\bf 121} 626-639. 
\bibitem{Jeff} Jeffreys H J 1924 On certain approximate solutions of linear differential equations of the second order. {\it Proc.  Lond. Math. Soc.} {\bf 23} 428-436.
\bibitem{FF} Fr\"{o}man H and Fr\"{o}man P O 1965 {\it JWKB Approximation - Contributions to the Theory} (North-Holland, Amsterdam).
\bibitem{prsa} Forbes R G and Deane J H B 2007 Reformulation of the Standard Theory of Fowler-Nordheim tunnelling and cold field electron emission {\it Proc. R. Soc. Lond. Ser. A.} accepted for publication (doi:10.1098/rspa.2007.0030). 
\bibitem{AS} Milne-Thompson L M 1965 Elliptic Integrals, in: {\it Handbook of Mathematical Functions} (eds. M Abramowitz and I A Stegun) (Dover, New York). 
\bibitem{BKH} Burgess R F, Kroemer H and  Houston J M 1953 Corrected values of Fowler-Nordheim field emission functions $v(y)$ and $s(y)$. {\it Phys. Rev.} {\bf 90} 515.
\bibitem{APL} Forbes R G 2006 Simple good approximations for the special elliptic functions in the JWKB theory of Fowler-Nordheim tunnelling through a Schottky-Nordheim barrier {\it Appl. Phys. Lett.} {\bf 89} 113122.
\bibitem{Hast} Hastings C, Jr 1955 {\it Approximations for Digital Computers} (Princeton University Press, Princeton, N.J.).
\bibitem{wylie} Wylie C R 1979, {\it Differential Equations} (McGraw-Hill Kogakusha: Tokyo).
\bibitem{Birk} Birkhoff G and Rota G-C {\it Ordinary Differential Equations} (2nd edition) (Blaisdell:
Waltham, Mass.). 
\bibitem{cayley} Cayley A 1876, {\it An Elementary Treatise on Elliptic Functions},
(Deighton Bell: Cambridge).
\bibitem{Chicago} Deane J H B and Forbes R G 2007 Exact mathematical solution for the principal field emission correction function $v$ used in Standard Fowler-Nordheim theory
{\it Technical Digest, 20th Intern. Vacuum Nanoelectronics Conf., Chicago, July 2007} (eds. H H Busta, C A Spindt, C E Hunt, I Brodie, E H Edwards) (ISBN: 1-4244-1134-3) (IEEE: Piscataway, NJ,) pp. 149-150.

\end{thebibliography}
\end{document}